\documentclass[fleqn,usenatbib]{mnras}

\usepackage{newtxtext,newtxmath}

\usepackage[T1]{fontenc}

\DeclareRobustCommand{\VAN}[3]{#2}
\let\VANthebibliography\thebibliography
\def\thebibliography{\DeclareRobustCommand{\VAN}[3]{##3}\VANthebibliography}

\usepackage{graphicx}	%
\usepackage{amsmath}	%
\usepackage{MnSymbol}	%
\usepackage{cancel} 	%

\newcommand{\xHI}{\langle x_\mathrm{HI} \rangle}
\newcommand{\NHI}{N_\mathrm{HI}^\mathrm{DW}}
\newcommand{\NHIabs}{N_\mathrm{HI}^\mathrm{abs}}

\newcommand{\Rb}{r_\mathrm{patch}}
\newcommand{\tQ}{t_\mathrm{Q}}
\newcommand{\logNHI}{\log_{10} \NHI/\mathrm{cm}^{-2}}

\newcommand{\logtQ}{\log_{10} \tQ/\mathrm{yr}}

\title[Local quasar IGM damping wing constraints]{First constraints on the local ionization topology in front of two quasars at $z \sim 7.5$}

\author[Kist et al.]{
Timo Kist,$^{1}$\thanks{E-mail: kist@strw.leidenuniv.nl}
Joseph F. Hennawi,$^{1,2}$
Frederick B. Davies,$^{3}$
Eduardo Ba\~nados,$^{3}$
Sarah E.~I.~Bosman,$^{3,4}$
\newauthor
Zheng Cai,$^{5}$
Anna-Christina Eilers,$^{6}$
Xiaohui Fan,$^{7}$
Zoltán Haiman,$^{8}$
Hyunsung D. Jun,$^{9}$
Yichen Liu,$^{7}$
\newauthor
Jinyi Yang$^{10}$ and
Feige Wang$^{10}$
\\
$^{1}$Leiden Observatory, Leiden University, P.O. Box 9513, 2300 RA Leiden,
The Netherlands\\
$^{2}$Department of Physics, University of California, Santa Barbara, CA 93106, USA\\
$^{3}$Max-Planck-Institut für Astronomie, Königstuhl 17, 69117 Heidelberg, Germany\\
$^{4}$Institute for Theoretical Physics, Heidelberg University, Philosophenweg 12, 69120, Heidelberg, Germany\\
$^{5}$Department of Astronomy, Tsinghua University, Beijing 100084, China\\
$^{6}$MIT Kavli Institute for Astrophysics and Space Research, 77 Massachusetts Avenue, Cambridge, MA 02139, USA\\
$^{7}$Steward Observatory, University of Arizona, 933 N Cherry Avenue, Tucson, AZ 85721, USA \\
$^{8}$Institute of Science and Technology Austria (ISTA), Am Campus 1, Klosterneuburg, Austria\\
$^{9}$Department of Physics, Northwestern College, 101 7th St. SW, Orange City, IA 51041, USA\\
$^{10}$Department of Astronomy, University of Michigan, 1085 S. University Ave., Ann Arbor, MI 48109, USA
}

\date{Accepted XXX. Received YYY; in original form ZZZ}

\pubyear{2025}

\begin{document}
\label{firstpage}
\pagerange{\pageref{firstpage}--\pageref{lastpage}}
\maketitle

\begin{abstract}
Thus far, Lyman-$\alpha$ damping wings towards quasars have been used to probe the \textit{global} ionization state of the foreground intergalactic medium (IGM). A new parameterization has demonstrated that the damping wing signature also carries \textit{local} information about the distribution of neutral hydrogen (HI) in front of the quasar before it started shining. Leveraging a recently introduced Bayesian \texttt{JAX}-based Hamiltonian Monte Carlo (HMC) inference framework, we derive constraints on the Lorentzian-weighted HI column density $\NHI$, the quasar's distance $\Rb$ to the first neutral patch and its lifetime $\tQ$ based on JWST/NIRSpec spectra of the two $z \sim 7.5$ quasars J1007+2115 and J1342+0928. After folding in model-dependent topology information, we find that J1007+2115 (and J1342+0928) is most likely to reside in a $\xHI = 0.32_{-0.20}^{+0.22}$ ($0.58_{-0.23}^{+0.23}$) neutral IGM while shining for a remarkably short lifetime of $\logtQ = 4.14_{-0.18}^{+0.74}$ (an intermediate lifetime of $5.64_{-0.43}^{+0.25}$) along a sightline with $\logNHI = 19.70_{-0.86}^{+0.35}$ ($20.24_{-0.22}^{+0.25}$) and $\Rb = 28.9_{-14.4}^{+54.0} \,\mathrm{cMpc}$ ($10.9_{-5.9}^{+5.6} \,\mathrm{cMpc}$). In light of the potential presence of local absorbers in the foreground of J1342+0928 as has been recently suggested, we also demonstrate how the Lorentzian-weighted column density $\NHI$ provides a natural means for quantifying their contribution to the observed damping wing signal.
\end{abstract}


\section{Introduction}

The highly sensitive Lyman-$\alpha$ transition observed in the spectra of high-redshift sources carries a wealth of information about the epoch of reionization, most particularly when the absorption imprint from the foreground neutral intergalactic medium (IGM) saturates, manifesting in a characteristic \textit{damping wing} signature redward of Lyman-$\alpha$ line center \citep{miralda-escude1998}. Quasars \citep[e.g.][]{mortlock2011, greig2017b, greig2019, greig2022, davies2018a, wang2020, yang2020, durovcikova2024} as well as galaxies \citep[e.g.][]{curtis-lake2023, mason2025} have been used as background sources to infer constraints on the global volume-averaged fraction $\xHI$ of neutral hydrogen (HI) at the redshift of the source. Stacking spectra at a similar redshift, or statistically combining the inferred $\xHI$ constraints makes IGM damping wings a natural probe of the timing of reionization. But recent studies have constructed parameterizations that capture the shape of the IGM damping wing significantly more tightly than the global IGM neutral fraction $\xHI$, and even carry information about the \textit{local} ionization topology in front of a given source \citep{chen2024, keating2024a, mason2025, kist2025b}.

Specifically, \citet{kist2025b} put forward a three-parameter model applicable in the context of quasars, consisting of two summary statistics of the $x_\mathrm{HI}$ field \textit{before} the quasar started shining (which we dub the \textit{pre}-quasar $x_\mathrm{HI}$ field), and the quasar's lifetime $\tQ$ encoding the effects of its ionizing radiation. The former two statistics of the pre-quasar ionization topology are 1) the column density $\NHI$ of neutral material in front of the quasar, weighted by a Lorentzian profile mimicking the frequency dependence of the Lyman-$\alpha$ cross section $\sigma_\alpha$ appearing in the optical depth integral; and 2) the distance $\Rb$ from the quasar to the first neutral patch. In a companion paper, \citet{kist2025c} introduced a fully Bayesian pipeline to infer these two parameters along with the quasar lifetime $\tQ$ in a reionization model-independent fashion based on observed high-redshift quasar spectra, accounting for all relevant sources of uncertainty such as the unknown intrinsic continuum of the quasar, IGM transmission fluctuations and spectral noise. Statistical tests on hundreds of mock spectra demonstrated that the pipeline allows them to constrain $\tQ$ to $0.58_{-0.13}^{+0.13}\,\mathrm{dex}$, $\NHI$ to $0.72_{-0.25}^{+0.08}\,\mathrm{dex}$ and $\Rb$ to $35.7_{-31.9}^{+7.8}\,\mathrm{cMpc}$ in case a noticeable damping wing is present in the spectrum. Here we take advantage of this framework to infer the first \textit{local} IGM damping wing constraints from JWST/NIRSpec spectra of two of the highest-redshift quasars known to date, that is, J1007+2115 at $z = 7.51$ \citep{yang2020}
and J1342+0928 at $z = 7.54$ \citep{banados2018}.

We briefly summarize in Section~\ref{sec:methods} the underlying theory and our local damping wing analysis framework, and proceed in Section~\ref{sec:data_inference} by introducing the data and presenting our analysis results for the two aforementioned objects. We put these measurements into context with existing literature constraints and conclude in Section~\ref{sec:conclusions}.

\section{Methods}
\label{sec:methods}

We conduct our analysis in the context of the local damping wing parameterization put forward by \citet{kist2025b}, harnessing the inference pipeline originally introduced in \citet{hennawi2025} and adapted by \citet{kist2025c} for the use with this parameterization. \citet{kist2025c} also conceptualized a theoretical framework for tying the inferred local parameter constraints which are agnostic to the underlying reionization topology to a specific reionization model to obtain a global constraint on the timing of reionization. We restrict this section to a brief but self-contained summary of each of these analysis components and refer the reader to the works above for more comprehensive descriptions of the respective parts.

The first of the two local summary statistics of the pre-quasar ionization topology defined by \citet{kist2025b} that we aim to constrain in this work is the Lorentzian-weighted HI column density
\begin{align}
\label{eq:NHI_DW}
    N_\mathrm{HI}^\mathrm{DW} \equiv\; &5.1 \times 10^{20}\,\mathrm{cm}^{-2}\times \left(\frac{r_\mathrm{T}}{18\,\mathrm{cMpc}}\right) \\ \nonumber
    &\times \left(\frac{1+z_\mathrm{QSO}}{1+7.54}\right)^2 \int_{r_\mathrm{min}/r_\mathrm{T}}^{r_\mathrm{max}/r_\mathrm{T}}  \frac{x_\mathrm{HI}(r) \cdot \Delta(r)}{(r/r_\mathrm{T} + 1)^2} \; \mathrm{d}(r/r_\mathrm{T}),
\end{align}
where $r$ denotes the comoving distance, and the denominator in the integrand is the Lorentzian weighting kernel which accounts for the frequency dependence of the Lyman-$\alpha$ cross section that the neutral hydrogen field $x_\mathrm{HI}(r)$ and the overdensity field $\Delta(r)$ are integrated against in the optical depth integral. The three distances $r_\mathrm{min}$, $r_\mathrm{max}$ and $r_\mathrm{T}$ are hyperparameters which we fix to $r_\mathrm{min} = 4\,\mathrm{cMpc}$, $r_\mathrm{max} = r_\mathrm{min} + 100\,\mathrm{cMpc}$ and $r_\mathrm{T} = 18\,\mathrm{cMpc}$ throughout \citep[for details see][]{kist2025b}. %
Our second statistic $\Rb$ can be straightforwardly defined as the (comoving) distance between the source and the first neutral patch. Recall that both $\NHI$ and $\Rb$ are summaries of the \textit{pre}-quasar ionization topology $x_\mathrm{HI}$ as this is the cosmological field we are aiming to constrain. Our third parameter, the lifetime $\tQ$ of the quasar, summarizes the effects of the quasar's ionizing radiation.

We simulate IGM transmission profiles $\boldsymbol{t}$ based on a hybrid approach following \citet{davies2018a} where we combine density, velocity and temperature skewers from the Nyx hydrodynamical simulations \citep{almgren2013, lukic2015} and synthetic $x_\mathrm{HI}$ skewers \citep{kist2025b} generated at $21$ column density values between $17.48 \leq \logNHI \leq 21.08$ and $18$ different neutral patch distances between $0.3\,\mathrm{cMpc} \leq \Rb \leq 143.0\,\mathrm{cMpc}$. We then perform one-dimensional radiative transfer \citep{davies2016} assuming $51$ quasar lifetimes between $10^3\,\mathrm{yr} \leq \tQ \leq 10^8\,\mathrm{yr}$, and forward-model realistic instrumental effects \citep{hennawi2025}. All our models are generated assuming $z_\mathrm{QSO} = 7.54$ with an ionizing photon emission rate of $Q = 10^{57.14}\,\mathrm{s}^{-1}$, explicitly resembling J1342+0928 but also appropriate for J1007+2115. Further, to tie our local constraints to the global IGM neutral fraction $\xHI$ in the way described in \citet{kist2025c}, we determine the stochastic mapping $P_\mathrm{top}(\logNHI, \Rb \,|\, \xHI)$,\footnote{Throughout, we use a 'top' subscript to denote probability distributions that are defined in the context of a specific reionization model.} by combining the same hydrodynamical Nyx sightlines with $x_\mathrm{HI}$ skewers extracted from semi-numerical reionization topologies simulated with a modified version of \texttt{21cmFAST} \citep{mesinger2011, davies2022} at $21$ different global IGM neutral fractions between $0 \leq \xHI \leq 1$.

In combination with a principal component analysis (PCA) model for the intrinsic quasar continuum $\boldsymbol{s}$ constructed based on a dataset of $44,587$ quasar spectra at $1.878 <
z < 3.427$  from the SDSS-III Baryon Oscillation Spectroscopic Survey (BOSS) and SDSS-IV Extended BOSS (eBOSS) \citep{hennawi2025b, kist2025c}, this allows us to write down the likelihood of an observed spectrum $\boldsymbol{f}$ given the astrophysical parameters $\boldsymbol{\theta}$, the PCA coefficients $\boldsymbol{\eta}$ and its observational noise vector $\boldsymbol{\sigma}$ as the following multivariate Gaussian distribution \citep{hennawi2025}:
\begin{equation}
\label{eq:likelihood}
    L(\boldsymbol{f}|\boldsymbol{\sigma}, \boldsymbol{\theta}, \boldsymbol{\eta}) = \mathcal{N}\left(\boldsymbol{f} ;\langle\boldsymbol{t}\rangle \circ\langle\boldsymbol{s}\rangle, \boldsymbol{\Sigma}+\langle\boldsymbol{S}\rangle \boldsymbol{C}_{\boldsymbol{t}}\langle\boldsymbol{S}\rangle+\langle\boldsymbol{T}\rangle \boldsymbol{C}_{\boldsymbol{s}}\langle\boldsymbol{T}\rangle\right),
\end{equation}
where $\boldsymbol{t}\circ\boldsymbol{s}$ is the element-wise (Hadamard) product of the two mean vectors $\langle\boldsymbol{t}\rangle$ and $\langle\boldsymbol{s}\rangle$, and  $\boldsymbol{C}_{\boldsymbol{t}}$ and $\boldsymbol{C}_{\boldsymbol{s}}$ are the covariance matrices of $\boldsymbol{t}$ and $\boldsymbol{s}$, and we write $\boldsymbol{\Sigma} \equiv \mathrm{diag}(\boldsymbol{\sigma})$,  $\boldsymbol{T}\equiv\mathrm{diag}(\boldsymbol{t})$ and $\boldsymbol{S}\equiv\mathrm{diag}(\boldsymbol{s})$.

We follow \citet{kist2025c} and initially infer the local parameters $\boldsymbol{\theta} \equiv (\logtQ, \logNHI, \Rb)$ under the assumption of a logarithmically flat prior on the lifetime between $3 \leq \logtQ \leq 8$, and a constant two-dimensional topology-agnostic prior on $(\logNHI, \Rb)$ with a non-trivial boundary enclosing all physically permitted regions of parameter space \citep[for details see][]{kist2025c}. After marginalizing out the nuisance parameters $\boldsymbol{\eta}$, we can probabilistically tie these local constraints on $\boldsymbol{\theta}$ to the aforementioned semi-numerical reionization topology via the conditional distribution $P_\mathrm{top}(\logNHI, \Rb \,|\, \xHI)$ and obtain a joint constraint on the global IGM neutral fraction:
\begin{align}
\label{eq:posterior_final}
    P_\mathrm{top}&(\xHI, \boldsymbol{\theta} | \boldsymbol{f}, \boldsymbol{\sigma}) \!=\! L(\boldsymbol{f} | \boldsymbol{\sigma}, \boldsymbol{\theta})\,\times\, P(\xHI) \times P(\logtQ) \nonumber \\
    &\times P_\mathrm{top}(\logNHI, \Rb \,|\, \xHI) / P(\boldsymbol{f} | \boldsymbol{\sigma}).
\end{align}
For comparison, we also perform the inference following the conventional approach without local summary statistics where $\boldsymbol{\theta} \equiv (\xHI, \logtQ)$ such that no conversion according to Eq.~(\ref{eq:posterior_final}) is necessary. We henceforth refer to this as the \textit{global} parameterization (as opposed to our \textit{local} one). Here we assume a flat neutral fraction prior between $0 \leq \xHI \leq 1$.

Practically, we sample from the respective posterior distribution via the \texttt{NumPyro} Hamiltonian Monte-Carlo (HMC) implementation with No U-Turn Sampler (NUTS). Each inference run consists of $8$ HMC chains with $1000$ warm-up and $2000$ sampling steps per chain \citep[for details, see][]{hennawi2025, kist2025c}. We reweight these samples based on the coverage tests performed in \citet{kist2025c} to ensure that we are quoting statistically faithful constraints.

\section{Local IGM damping wing constraints at $\MakeLowercase{z} \sim 7.5$}
\label{sec:data_inference}

\begin{figure*}
	\includegraphics[width=\textwidth]{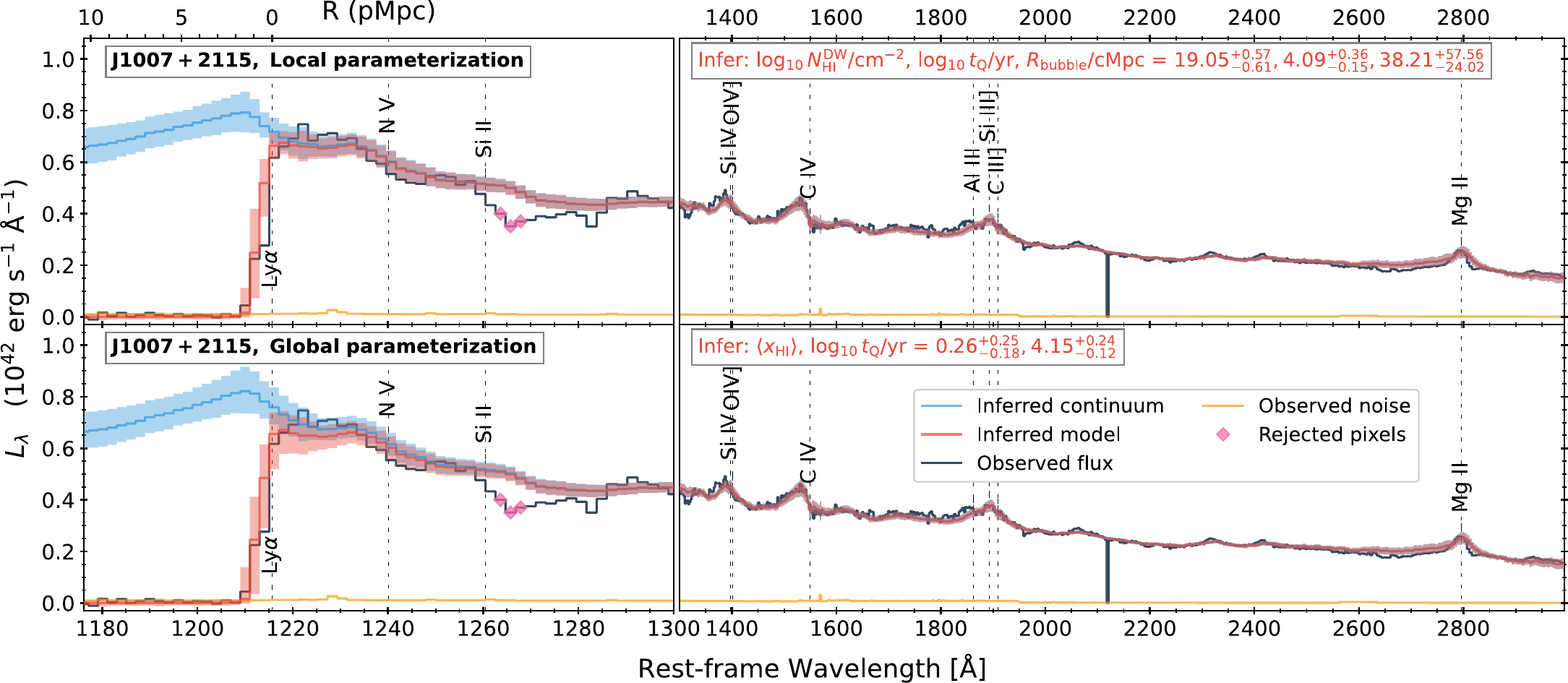}
    \caption{Inferred model for the JWST/NIRSpec spectrum of the quasar J1007+2115, fitted in the context of the local IGM damping wing parameterization (upper row) and the global one (lower row). The observed (and rebinned) spectrum of the quasar is depicted in black, with the noise vector shown in yellow. The inferred model spectrum is depicted in red and the unabsorbed inferred continuum in blue, where solid lines represent the median inferred models, and shaded regions the $16\,\%$ and the $84\,\%$ percentile variations reflecting parameter uncertainty, continuum reconstruction errors, as well as spectral noise.}
    \label{fig:spec_J1007}
\end{figure*}

\begin{figure}
	\includegraphics[width=\columnwidth]{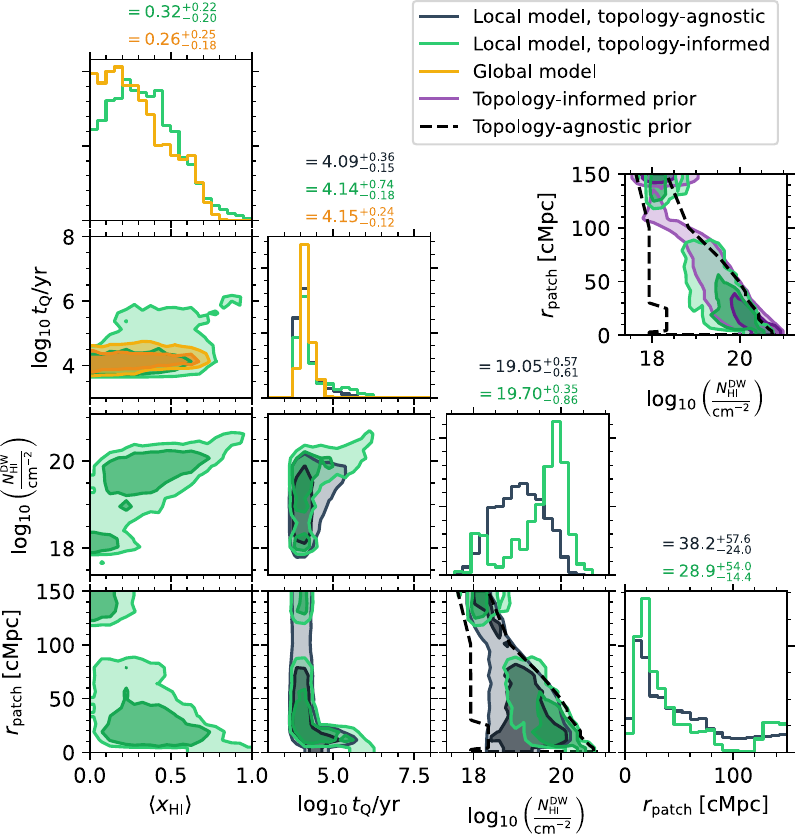}
    \caption{Posterior distributions inferred from the JWST/NIRSpec spectrum of the quasar J1007+2115 depicted in Figure~\ref{fig:spec_J1007} in the context of the local IGM damping wing parameterization (black) and the global one (yellow). Both distributions are marginalized over $7$ nuisance parameters describing the shape of the quasar continuum. Additionally depicted in green is the topology-informed version of the local constraints, entailing the non-trivial prior $P_\mathrm{top}(\logNHI, \Rb)$ (explicitly depicted in purple in the extra panel), and also providing a constraint on the global IGM neutral fraction $\xHI$ in good agreement with the directly inferred one.}
    \label{fig:corner_J1007}
\end{figure}

\begin{figure*}
	\includegraphics[width=\textwidth]{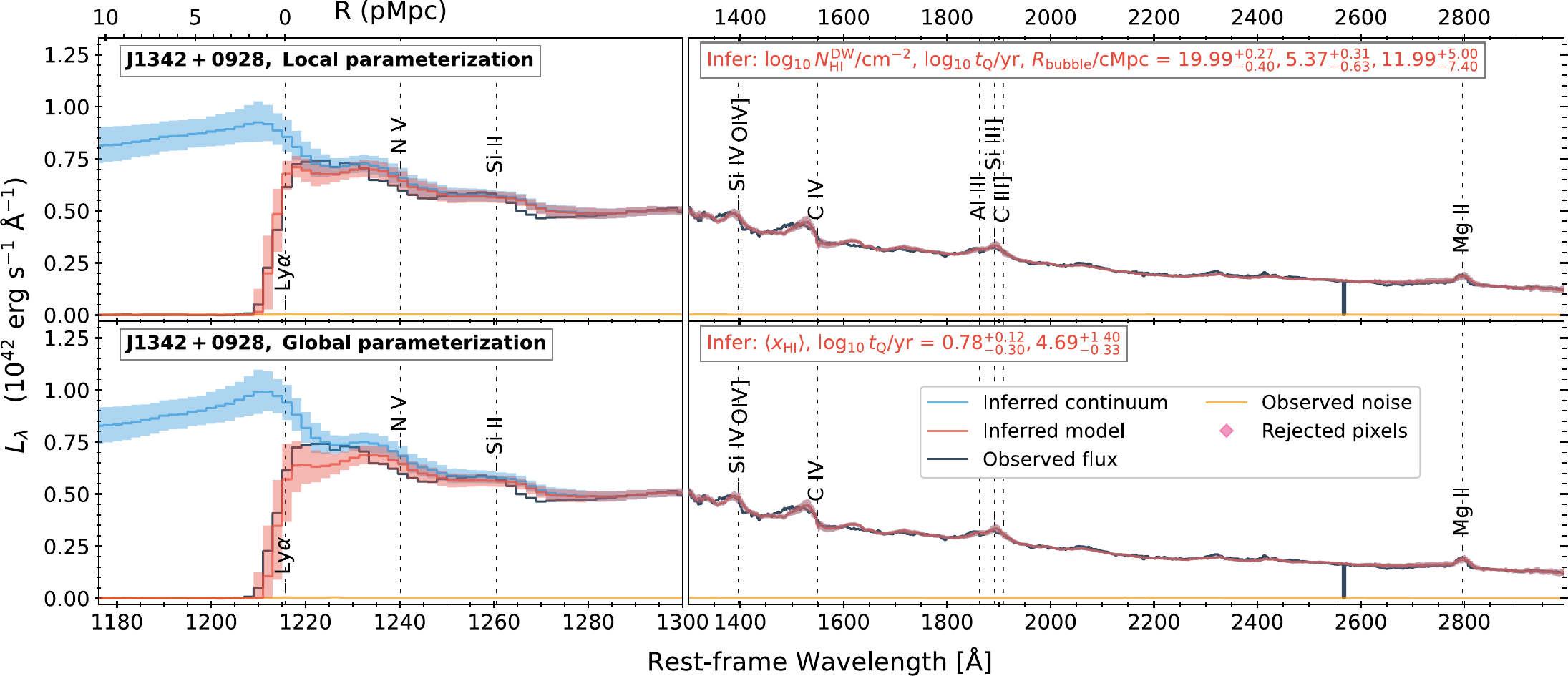}
    \caption{Inferred models like in Figure~\ref{fig:spec_J1007} but for the JWST/NIRSpec spectrum of the quasar J1342+0928.}
    \label{fig:spec_J1342}
\end{figure*}

\begin{figure}
	\includegraphics[width=\columnwidth]{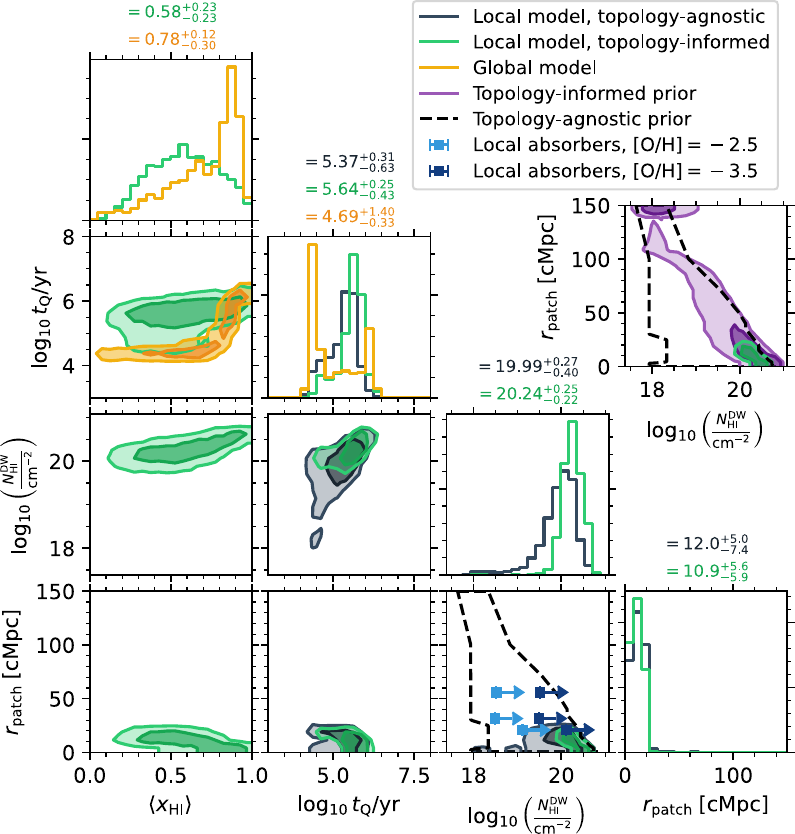}
    \caption{Like Figure~\ref{fig:corner_J1007} but for the posterior distributions inferred from the JWST/NIRSpec spectrum of the quasar J1342+0928 depicted in Figure~\ref{fig:spec_J1342}. The $(\NHI, \Rb)$ panel also shows the putative local absorber constraints obtained by \citet{davies2025} assuming two different metallicities which provide lower limits on $\NHI$.}
    \label{fig:corner_J1342}
\end{figure}

We now proceed by leveraging the framework summarized in the previous section to infer the first \textit{local} IGM damping wing constraints for two of the highest-redshift quasars known to date. That is, J1007+2115 at $z = 7.51$ \citep{yang2020}
and J1342+0928 at $z = 7.54$ \citep{banados2018} with absolute magnitudes of $M_{1450} = -26.82$ and $M_{1450} = -26.34$ at $1450\,\text{\AA}$ in the rest frame, respectively, determined based on Euclid photometry \citep{yang2025}.

This work presents the first damping wing analysis of JWST/NIRSpec spectra of these two objects \citep[JWST GO 1764, PI: Fan, and JWST GTO 1219, PI: Luetzgendorf; see also][]{christensen2023, hennawi2025b}. The spectra have signal-to-noise ratios of $\mathrm{S}/\mathrm{N} \simeq 5 - 20$ at a resolution of $R \simeq 2700$ and were taken with the NIRSpec Fixed Slits together with the G140H/F070LP and G235H/F170LP grating and filter combinations. The data were reduced via a combination of the JWST Science calibration pipeline \texttt{CALWEBB} (version 1.13.4) and the \texttt{python}-based semi-automated reduction pipeline \texttt{PypeIt} \citep{prochaska2020}. The full details of the reduction procedure are discussed in \citet{hennawi2025b}.

The spectral coverage of these observations allows us to operate on a rest-frame wavelength range of $1175\,\text{\AA} \leq \lambda \leq 3000\,\text{\AA}$. After removing narrow absorbers, we rebin both spectra to a velocity pixel scale of $\mathrm{d}v = 500\,\mathrm{km}/\mathrm{s}$ (and likewise are our forward-modeled IGM transmission profiles) as our current likelihood prescription does not allow us to extract all information from the full resolution data at its native pixel scale in a statistically faithful manner \citep{kist2025a, kist2025c} which is pending improvements in our likelihood prescription, for example through simulation-based inference \citep{chen2024}.

\subsection{J1007+2115}
\label{sec:J1007}

We start by analyzing the quasar J1007+2115 at $z = 7.51$. Previous studies have inferred IGM neutral fractions of $\xHI = 0.39_{-0.13}^{+0.22}$ \citep{yang2020} and $\xHI = 0.27_{-0.17}^{+0.21}$ \citep{greig2022} from ground-based spectra of this object. Note that the former analysis also obtained a constraint of $\logtQ = 4.89_{-0.76}^{+1.15}$ on its lifetime \citep{eilers2021}, and used, upon some minor differences, the same simulation models as we do in this work, however with a significantly different analysis pipeline.
Our re-analysis is based on the new JWST/NIRSpec spectrum of this object, %
and we perform the analysis both in the context of the conventional global as well as our local IGM damping wing parameterization. We depict the final rebinned spectrum that forms the input to our pipeline as the black line in Figure~\ref{fig:spec_J1007} with associated noise vector in yellow. The upper panel depicts our reconstruction in the context of the local parameterization, and the lower panel in the context of the global one. The median reconstructed continua are shown in blue, and the full models including IGM absorption in red. The shading around these lines highlights the associated $16\,\%$ and $84\,\%$ percentile variations, reflecting parameter uncertainty, continuum reconstruction errors, as well as spectral noise \citep[for further details see][]{hennawi2025}.

Overall, the models show a remarkable agreement, both providing a good fit to the spectrum across the entire spectral range, not only matching the shape of the proximity zone and the Lyman-$\alpha$ damping wing, but also the smooth emission lines redward of Lyman-$\alpha$. The median models slightly undershoot the observed spectrum around $\lambda \approx 1225\,\text{\AA}$ (somewhat more pronounced in the global parameterization), and slightly overshoot at $\lambda \approx 1245\,\text{\AA}$. However, in both regions the observed value is still within the $68\,\%$ region. Somewhat more redward, at $\lambda \approx 1265\,\text{\AA}$, three pixels got rejected by the sigma-clipping procedure we apply prior to the inference \citep[see][]{kist2025c}. These pixels are possibly affected by a mild broad absorption line (BAL) system but this does not appear to impact our conclusions as we obtain the same results when no clipping is applied.

The only main difference between the global and the local model curves is a reduced degree of scatter 
in the local parameterization as this parameterization excludes the scatter due to the stochastic distribution of neutral patches along the line-of-sight
from the inference task. Both models only show a very mild damping wing. The black contours in Figure~\ref{fig:corner_J1007} depict the associated local parameter constraints. The lifetime posterior peaks relatively sharply at $\logtQ = 4.09_{-0.15}^{+0.36}$, implying that we are concerned with a comparably young object, as already hinted at by its small proximity zone \citep[see e.g. also the objects in][]{eilers2020, eilers2021}. 
The data are not overly constraining with respect to the two local parameters $\NHI$ and $\Rb$. In essence, only the highest HI column densities (and correspondingly short neutral bubble distances that are still located in the physically permitted region of parameter space enclosed by the dashed line in the $(\NHI, \Rb)$ panel) are excluded as these would lead to a noticeable damping wing imprint.

We obtain a global $\xHI$ constraint on the timing of reionization based on these local constraints by following the procedure introduced in \citet{kist2025c} and summarized in Section~\ref{sec:methods}, folding in the distribution of our local parameters within the realistic semi-numerical reionization topology as a prior. 
The green contours in Figure~\ref{fig:corner_J1007} show the resulting posterior distribution. We see that the prior on $\NHI$ and $\Rb$ induced by this topology (shown explicitly in purple in the extra panel of the plot) disfavors the (low-$\NHI$, low-$\Rb$) region of the posterior and therefore gives more weight to the (low-$\NHI$, high-$\Rb$) and (high-$\NHI$, low-$\Rb$) peaks.\footnote{Note that the green converted posterior seemingly has support in regions where the black unconverted one does not. In fact, \textit{both} posteriors have support in these regions, but before performing the conversion, these are not included in the $95\,\%$ contour shown in the plot. Further, due to smoothing effects and the statistical reweighting procedure we apply, the contours marginally extend beyond the physical $(\NHI, \Rb)$ boundary enclosed by the dashed line.} The resulting $\xHI$ posterior suggests a comparably low to intermediate neutral fraction of $\xHI = 0.32_{-0.20}^{+0.22}$ at this redshift. The lifetime constraint largely remains unaffected by folding in the topology information, still peaking at $\logtQ = 4.14_{-0.18}^{+0.74}$, on the lower end of the literature value of $\logtQ = 4.89_{-0.76}^{+1.15}$ \citep{yang2020, eilers2021}. Our local parameter constraints, on the other hand, are fully prior-dominated, as a comparison of the green and purple contours in the extra $(\NHI, \Rb)$ panel in the same figure suggests.

We test the robustness of the $\xHI$ constraint by also comparing it to the one we obtain when inferring $\xHI$ and $\tQ$ directly in the context of the global parameterization. The corresponding model curves are shown in the lower panel of Figure~\ref{fig:spec_J1007}. The resulting posterior, depicted by the yellow contours in Figure~\ref{fig:corner_J1007}, is in good agreement with the one obtained by converting our local constraints (green). The only differences 
between the two are that the converted posterior exhibits a somewhat more extended tail towards longer lifetimes. Since these longer lifetimes go hand in hand with a higher IGM neutral fraction, the marginal $\xHI$ posterior obtained via the local parameterization (green)
shows a very mild peak at around $\xHI \simeq 0.3$, in contrast to the directly inferred one which peaks very close to $\xHI \simeq 0$. Overall, however, both posteriors are in clear statistical agreement, underlining the robustness of our constraints despite the highly different ways in which they were obtained.

\subsection{J1342+0928}
\label{sec:J1342}

The second object of our study, the quasar J1342+0928 at $z = 7.54$, has already been extensively analyzed in previous literature \citep{banados2018, davies2018a, greig2019, durovcikova2020, reiman2020, greig2022}, resulting in varying constraints on the IGM neutral fraction, with median values ranging from $\xHI \approx 0.2 - 0.6$. We again apply both versions of our inference framework to the JWST/NIRSpec spectrum of this object and obtain the fits and parameter constraints shown in Figures~\ref{fig:spec_J1342} and \ref{fig:corner_J1342}, again in good statistical agreement with a reduced degree of scatter in the local parameterization. Unlike in the case of J1007+2115, we infer the clear presence of a damping wing with a correspondingly high HI column density of $\logNHI = 19.99_{-0.40}^{+0.27}$ and a rather short distance of $\Rb = 12.0_{-7.4}^{+5.0} \,\mathrm{cMpc}$ to the first neutral patch, increasing/reducing further to $\logNHI = 20.24_{-0.22}^{+0.25}$ and $\Rb = 10.9_{-5.9}^{+5.6}$ after folding in the topology dependence (green contours). The lifetime posterior shows a clear degeneracy with $\NHI$ in the local parameterization, peaking at $\logtQ = 5.64_{-0.43}^{+0.25}$, and with $\xHI$ in the global one, preferring a somewhat lower value of $\logtQ = 4.69_{-0.33}^{+1.40}$. However, the global posterior shows a long axis of degeneracy with a second peak closer to the higher value preferred in the local model. Previous analyses inferred $\logtQ = 5.38_{-1.30}^{+0.72}$ with a similarly wide degeneracy \citep{davies2018a, davies2019, eilers2021}.
The corresponding $\xHI$ constraints suggest an intermediate to high IGM neutral fraction of $\xHI = 0.58_{-0.23}^{+0.23}$ in the context of the local parameterization, and a very high one of $\xHI = 0.78_{-0.30}^{+0.12}$ in the context of the global model.

These results ought to be treated with caution since \citet{davies2025} recently pointed out the possibility of contamination of the IGM in front of J1342+0928 by proximate
DLA absorbers located at $r_\mathrm{abs} \simeq 21\,\mathrm{cMpc}$, $31\,\mathrm{cMpc}$ and/or $56\,\mathrm{cMpc}$ based on weak Mg II absorption lines they identified in the JWST/NIRSpec spectrum that is also the subject of this analysis. While the authors
concluded that such proximate absorbers would have to be unusually
metal-poor, we use this possibility to highlight an additional virtue of our local parameterization: including the (Lorentzian-weighted) HI column density $\NHI$ as an explicit model parameter allows for a straightforward comparison of our constraints to the (unweighted) column density $\NHIabs$ of a putative absorber. The Lorentzian-weighted version $N_\mathrm{HI}^{\mathrm{DW, abs}}$ of the absorber's column density then combines with the column density $N_\mathrm{HI}^{\mathrm{DW, IGM}}$ sourced by the neutral IGM to the total column density of
\begin{equation}
    \NHI = N_\mathrm{HI}^{\mathrm{DW, IGM}} + N_\mathrm{HI}^{\mathrm{DW, abs}}
\end{equation}
giving rise to the practically observed damping wing signal. By approximating the localized absorber as a Dirac delta peak at position $r_\mathrm{abs}$, we can straightforwardly use Eq.~(\ref{eq:NHI_DW}) to determine $N_\mathrm{HI}^{\mathrm{DW, abs}}$ based on its classical column density $N_\mathrm{HI}^{\mathrm{abs}}$:
\begin{equation}
    \log N_\mathrm{HI}^{\mathrm{DW, abs}} = \log N_\mathrm{HI}^{\mathrm{abs}} - 2 \,\log (r_\mathrm{abs}/r_\mathrm{T} + 1),
\end{equation}
provided the absorber is located within the integration range for $\NHI$ enclosed by $r_\mathrm{min}$ and $r_\mathrm{max}$. Any inferred value of $\NHI$ therefore only constitutes an upper limit on the column density $N_\mathrm{HI}^{\mathrm{DW, IGM}}$ sourced by the IGM (and thus relevant to reionization), from which the contributions from any potential absorber would have to be subtracted. Vice versa, the column density $N_\mathrm{HI}^{\mathrm{DW, abs}}$ of a given absorber places a lower limit on the full $\NHI$. We plot these limits based on the values determined by \citet{davies2025} assuming oxygen abundances relative to solar of $[\mathrm{O}/\mathrm{H}] = -2.5$ ($-3.5$) as light (dark) blue arrows in the $(\NHI, \Rb)$ panel of Figure~\ref{fig:corner_J1342}. Note that the length of the arrows carries no information and we place these constraints at $\Rb = r_\mathrm{abs}$ for each absorber even though the two distances are not necessarily related given that we define $\Rb$ as the first \textit{extended} neutral patch in front of the quasar. The fact that the constraint from the closest absorber in the $[\mathrm{O}/\mathrm{H}] = -3.5$ case aligns well with the peak of the posterior is in excellent agreement with the fact that a such metal-poor absorber at this distance would solely account for the observed damping wing imprint without any additional IGM contribution. The Lorentzian-weighted HI column density of an $[\mathrm{O}/\mathrm{H}] = -2.5$ absorber, on the other hand, would largely be negligible compared to our inferred $\NHI$, and hence not bias our conclusions about the ionization state of the IGM. Our local parameterization thus also paves the way for a principled, \textit{joint} inference of the damping wing signal due to the IGM \textit{and} potential local absorbers which we leave to future work.

\section{Summary and Conclusions}
\label{sec:conclusions}

\begin{figure}
	\includegraphics[width=\columnwidth]{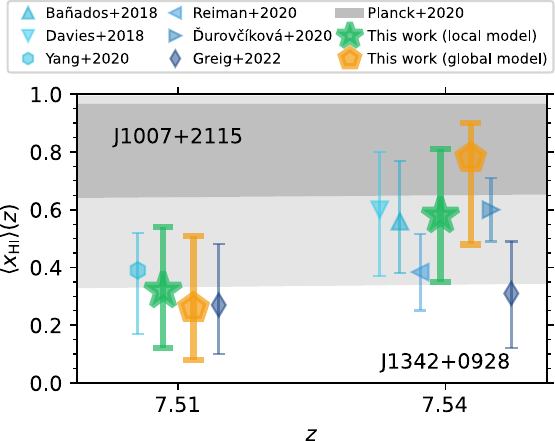}
    \caption{Reionization history constraints obtained from the two quasars J1007+2115 at $z=7.51$ and J1342+0928 at $z=7.54$ in the context of the local IGM damping wing parameterization (green stars) and the global one (yellow pentagons). Other symbols show existing literature constraints for the same two objects, slightly offset with respect to each other for clarity, and gray swathes are $68\,\%$ and $95\,\%$ contours of the \citet{planck_collaboration2020} CMB constraints.}%
    \label{fig:reion_hist}
\end{figure}

We summarize in Figure~\ref{fig:reion_hist} our constraints on the reionization history obtained by analyzing the JWST/NIRSpec spectra of the two $z \sim 7.5$ quasars J1007+2115 and J1342+0928 in the context of the local damping wing parameterization (green stars) put forward in \citet{kist2025b, kist2025c} and the conventional global one (yellow pentagons). We compare these constraints to the CMB constraint on the reionization optical depth inferred by the \citet{planck_collaboration2020} whose $68\,\%$- and $95\,\%$-contours are marked as gray swathes, as well as previous IGM damping wing measurements of the two objects studied in this work \citep{banados2018, davies2018a, yang2020, reiman2020, durovcikova2020, greig2022}. 
Our constraints are in clear statistical agreement with all these literature constraints, also owing to the comparably large statistical uncertainties that are unavoidable for individual objects due to the stochastic nature of reionization \citep{kist2025a, kist2025b, kist2025c}. Note that despite the methodological improvements in our pipeline as compared to previous approaches \citep{hennawi2025, kist2025c}, our uncertainties are not necessarily smaller than those quoted in the literature. This is explained by the fact that we rigorously folded in all relevant uncertainties due to stochastic reionization, the unknown quasar lifetime, continuum reconstruction, and spectral noise.

We conclude that J1007+2115 prefers a somewhat lower neutral fraction of $\xHI = 0.32_{-0.20}^{+0.22}$, compared to $\xHI = 0.58_{-0.23}^{+0.23}$ based on J1342+0928 (both inferred in the framework of our local parameterization). The directly inferred global $\xHI$ posterior of the latter object ($\xHI = 0.78_{-0.30}^{+0.12}$) peaks at the high end compared to our local and most literature damping wing constraints (though closer to \citet{planck_collaboration2020}), but note here the extended axis of degeneracy of the full posterior of this object (c.f. Figure~\ref{fig:corner_J1342}). In addition, \textit{all} damping wing-based $\xHI$ constraints would be biased high in case an unusually metal-poor absorber was indeed present in the foreground of this object \citep{davies2025} which would pull the actual corrected IGM neutral fraction closer to the value inferred from J1007+2115. Our local parameterization provides a natural framework for studying this possibility more carefully in future work, particularly relevant also to JWST observations of IGM damping wings towards galaxies \citep{mason2025, huberty2025}. More conclusive statements about the ionization state of the IGM at $z \sim 7.5$ will be enabled by applying our robust inference approach to the spectra of additional objects identified at these redshifts by the Euclid wide field survey \citep{euclid_collaboration2019, banados2025, yang2025}.

In addition, we constrained the lifetimes of the two objects of this study and found J1007+2115 to be remarkably young with $\logtQ = 4.14_{-0.18}^{+0.74}$, whereas J1342+0928 (with $\logtQ = 5.64_{-0.43}^{+0.25}$) is closer to the average expected lifetimes of $\tQ \sim 10^6\,\mathrm{yr}$ \citep{khrykin2021, morey2021}. Most importantly, we also obtained the first constraints on the local ionization topology in front of these two sources. With a Lorentzian-weighted HI column density of $\logNHI = 19.70_{-0.86}^{+0.35}$ and a first neutral patch at $\Rb = 28.9_{-14.4}^{+54.0} \,\mathrm{cMpc}$, we remained in the prior-dominated regime for the pre-quasar sightline originating from J1007+2115, but we measured $\logNHI = 20.24_{-0.22}^{+0.25}$ and $\Rb = 10.9_{-5.9}^{+5.6} \,\mathrm{cMpc}$ for J1342+0928. While this work was focused on the conversion of these measurements to constraints on the global timing on reionization, constraining $\NHI$ and $\Rb$ for larger statistical samples of objects will also open the doors to direct constraints on the topology of reionization with quasar IGM damping wings \citep{sharma2025}.

\section*{Acknowledgements}

We acknowledge helpful conversations with the ENIGMA group at UC Santa Barbara and Leiden University. This work made use of \texttt{NumPy} \citep{harris2020},  \texttt{SciPy} \citep{virtanen2020}, \texttt{JAX} \citep{bradbury2018}, \texttt{NumPyro} \citep{bingham2018, phan2019}, \texttt{sklearn} \citep{pedregosa2011}, \texttt{Astropy} \citep{astropy_collaboration2013, astropy_collaboration2018, astropy_collaboration2022}, \texttt{PypeIt} \citep{prochaska2020}, \texttt{SkyCalc\_ipy} \citep{leschinski2021}, \texttt{h5py} \citep{collette2013}, \texttt{Matplotlib} \citep{hunter2007}, \texttt{corner.py} \citep{foreman-mackey2016}, and \texttt{IPython} \citep{perez2007}.
TK and JFH acknowledge support from the European Research Council (ERC) under the European Union’s Horizon 2020 research and innovation program (grant agreement No 885301). JFH acknowledges support from NSF grant No. 2307180. SEIB is supported by the Deutsche Forschungsgemeinschaft (DFG) under Emmy Noether grant number BO 5771/1-1. FW acknowledges support from NSF award AST-2513040.

\section*{Data Availability}

The derived data generated in this research will be shared on reasonable requests to the corresponding author.

\bibliographystyle{mnras}
\bibliography{main} %

\appendix

\bsp	%
\label{lastpage}
\end{document}